\documentclass[12pt,preprint]{aastex}

\def\gsim{\,\lower3pt\hbox{$\sim$}\llap{\raise2pt\hbox{$>$}}\,}
\def\lsim{\,\lower3pt\hbox{$\sim$}\llap{\raise2pt\hbox{$<$}}\,}
\newcommand{\be}{\begin{equation}}
\newcommand{\ee}{\end{equation}}
\newcommand{\bex}{\begin{equation}\notag}
\newcommand{\eex}{\end{equation}\notag}
\newcommand{\bea}{\begin{eqnarray}}
\newcommand{\eea}{\end{eqnarray}}
\newcommand{\beax}{\begin{eqnarray*}}
\newcommand{\eeax}{\end{eqnarray*}}
\newcommand{\ba}{\begin{array}}
\newcommand{\ea}{\end{array}}

\begin{document}

\title{Emergence of Anchored Flux Tubes Through the Convection Zone}
\shorttitle{Emergence of Anchored Flux Tubes}


\author{G.~H. Fisher, D.-Y. Chou, A.~N. McClymont}
\affil{Institute for Astronomy, University of Hawaii,
Honolulu, HI 96822}


\begin{abstract}
We model the evolution of buoyant magnetic flux tubes in 
the Sun's convection
zone. A flux tube is assumed to lie initially near the top of the stably 
stratified radiative core below the convection zone, but a segment of it is
perturbed into the convection zone by gradual heating and convective overshoot
motions.  The ends (``footpoints'') of the segment remain anchored at the
base of the convection zone, and if the segment is sufficiently long, it may
be buoyantly unstable, rising through the convection zone in a short time.  The
length of the flux tube determines the ratio of buoyancy to magnetic tension:
short loops of flux are arrested before reaching the top of the convection zone,
while longer loops emerge to erupt through the photosphere.  Using Spruit's
convection zone model, we compute the minimum footpoint separation $L_c$
required for erupting flux tubes.  We explore the dependence of $L_c$ on the
initial thermal state of the perturbed flux tube segment and on its initial 
magnetic field strength.  Following an investigation of thermal diffusion
time scales and the dynamic rise times of unstable flux tube segments, we
conclude that the most likely origin for magnetic flux which erupts to the
surface is from short length scale perturbations ($L < L_c$) which are initially
stable, but which are subsequently destabilized either by diffusion of heat
into the tube or by stretching of the anchor points until $L$ 
just exceeds $L_c$.  In either case, the separation of the anchor points of the emergent tube should lie between the critical distance for a tube in mechanical
equilibrium and one in thermal equilibrium.  Finally, after comparing the
dispersion of dynamic rise times with the much shorter observed active region
formation time scales, we conclude that active regions form from the emergence
of a single flux tube segment.
\end{abstract}

\section{Introduction}
\label{section:intro}

In several recent theories of solar and stellar dynamos 
(\citealt{Golub1981}; \citealt{Galloway1981}; \citealt{DeLuca1986})
magnetic fields are generated
not in the convection zone, but in a thin, convectively stable layer just below
it.  In order for this magnetic flux to emerge to the photosphere where it can
be observed, it must first make its way through the convection zone.  It is
generally believed that magnetic buoyancy provides the force which pushes the
magnetic flux toward the surface 
\citep{Parker1979}.
Our goal in this paper is
to use the thin flux tube approximation of 
\cite{Spruit1981}
to study the emergence
of buoyant loops of magnetic flux whose ends are still anchored in the stable
layer below the convection zone.  Our approach is similar to that used by
\cite{Moreno-Insertis1986}.
Further details of our calculations may be found in
\cite{Chou1989}; henceforth Paper I, and
\cite{Fisher1989p}, henceforth Paper II.
Here we present only the essentials.

\section{Description of the Model}
\label{section:modeldescription}

In the thin flux tube approximation, the tube diameter is assumed to be smaller
than any other relevant physical length scale, and the magnetic field is taken
to be constant across the tube.  Furthermore, since the magneto-acoustic
transit time across the tube will be smaller than other time scales of interest,
one may assume that there is always a balance between the gas pressure outside
the tube and the sum of gas plus magnetic pressure inside the tube.  One can
think of a thin flux tube as a one-dimensional curve embedded in space.  Our
flux tube is tied down at both ends to the bottom of the convection zone, with
the portion in between free to move in response to buoyancy, magnetic tension,
and drag forces.  We take the flux tube to lie in a vertical plane,
defined by a horizontal $x-$ axis at the base of the convection zone, and a 
vertical $y-$ axis, and regard the height $y$ of the the flux tube as a function
of $x$ and time $t$.  The ends of the flux tube segment are anchored at $x=0$
and $x=L$.  In all the calculations discussed in this paper, we assume that
the flux tube initially ($t=0$) lies flat at the bottom of the convection zone
($y=0$).  In the discussion below, subscript $e$ refers to the external plasma
of the convection zone, and subscript $i$ refers to the plasma inside the flux
tube.

\subsection{The Basic Equations}
\label{subsection:equations}

The velocity of the plasma in the flux tube consists of the speed of the tube
itself in the direction normal to its length (denoted $v_n$) and the speed of
plasma moving along the flux tube (denoted by $v_s$).  The equation of motion
for the velocity component $v_n$ is given by
\be
(\rho_e + \rho_i) {d v_n \over d t} = {B^2 \over 4 \pi R} + 
{(\rho_e - \rho_i) g \over \sqrt{1 + ( \partial y / \partial x )^2}}
-{ C_D \rho_e | v_n | v_n \over \sqrt{\pi A}}
\label{equation:eqnmo}
\ee
where the terms on the right hand side of equation (\ref{equation:eqnmo}) 
correspond, respectively, to magnetic tension, buoyancy, and aerodynamic
drag forces.  The drag coefficient $C_D$ is taken to be unity.
Note that the inertial term contains the ``added mass'' of the displaced
external plasma flowing around the flux tube (see, e.g., 
\citealt{Landau1959},
\S 24).  An insignicant centrifugal force due to parallel flows through
bends in the flux tube has been neglected.  Equation (\ref{equation:eqnmo}) must
be supplemented by an equation relating the vertical motion of the tube to the
velocity $v_n$ and the flux tube slope $\partial y / \partial x$.  From
geometrical considerations, we find at a fixed value of $x$
\be
{ \partial y \over \partial t} = v_n \sqrt{1 + (\partial y / \partial x)^2}
\label{equation:geometry}\ .
\ee
The radius of curvature $R$, the magnetic field strength $B$, and the flux
tube cross-sectional area $A$ appearing in equation (\ref{equation:eqnmo}) are
determined by
\be
{1 \over R} = {\partial^2y / \partial x^2 \over 
[1 + ( \partial y / \partial x)^2 ]^{3/2}} ;\ P_e - P_i = {B^2 \over 8 \pi} ;\
B A = \Phi\ ,
\label{equation:aux}
\ee
where $P_i$ and $P_e$ are the internal and external gas pressures, respectively,
and $\Phi$ is the total magnetic flux.  The numerical techniques used for
solving equations (\ref{equation:eqnmo}) and (\ref{equation:geometry}) are
described in Paper I.

The parallel flows of the plasma inside the flux tube can be calculated from
equations (\ref{equation:eqnmo}) and (\ref{equation:geometry}) and mass 
conservation arguments (see Paper I).  These flows turn out to be sufficiently
slow that the plasma can be taken to be in hydrostatic equilibrium.  If we make
the same assumption regarding the external plasma (i.e., ignoring effects of
convection), then it is straightforward to determine all the thermodynamic and
magnetic variables as functions of height alone.  This procedure is outlined
below.  It is not at all obvious {\it a priori} that the neglect of convective
motions is justified.  The presence of these motions could affect flux tube
evolution in two ways.  First, the convective motions could severely distort
the shape of the flux tube.  Second, convective motions introduce a turbulent
pressure which perturbs hydrostatic equilibrium and this could affect the
calculated magnetic field strength.  We consider both effects in Paper I and
conclude that the first effect could be important for some of the thinnest
flux tubes we study, whereas the second is not important for any of the cases
we have studied.

The structure of the background external atmosphere is taken to be the 
convection zone model of 
\cite{Spruit1974}.
This atmosphere is slightly
superadiabatic due to convective heat transport.  The information in
Spruit's paper is sufficient to calculate the thermodynamic and magnetic
quantities needed for our dynamic model.  We assume that the plasma inside the
flux tube is isolated from heat transfer and behaves adiabatically.  There will
of course be heat transport between the flux tube and its surroundings, but we
have not included this specifically in the model.  Our approach in this paper
is to approximate the effects of heat conduction by choosing a physically
self-consistent value of the temperature difference between the flux tube and
its environs ($i.e.$ the parameter $\eta$ introduced below) at the base of the
convection zone.  This will be described further in the section on time scales
for emerging flux tubes.

It is straightforward to derive an equation for $\beta$ 
[$\beta \equiv 8 \pi P_e / B^2 = P_e / ((P_e-P_i)$] in terms of the height
variation of the external temperature difference $\delta T \equiv T_e-T_i$
between the external and internal plasma; this avoids the numerical difficulty
of subtracting two gas pressures which are nearly equal.  Starting from the
equilibrium equations
\be
{d P_e \over d y} = - \rho_e g ;\  {d P_i \over d y} = -\rho_i g
\label{equation:hse}
\ee
one finds
\be
{1 \over \beta ( \beta - 1)} {d \beta \over d y} = - {1 \over \Lambda}
{ \delta T / T_e \over 1 - \delta T / T_e}
\label{equation:beta}
\ee
where $\Lambda$ is the external pressure scale height.  This has the solution
\be
\beta (y)^{-1} = \beta_0^{-1} \exp [ - \gamma (y) ] + 
\{ 1 - \exp [ - \gamma (y) ] \} 
\label{equation:betasoln}
\ee
where $\beta_0 = \beta (0)$ and
\be
\gamma (y) = \int_0^y { dy' \over \Lambda(y') } {\delta T (y') / T_e (y') \over
1 - \delta T(y') / T_e(y') }\ .
\label{equation:gamma}
\ee
The temperature difference is simply
\be
\delta T(y) = \delta T_0 - \int_0^y dy' ( \nabla - \nabla_{ad} ) \mu g / R_g
\label{equation:deltat}
\ee
where $\delta T_0 = \delta T(0)$, 
$\nabla \equiv ( \partial ln T / \partial ln P )$,
$\nabla_{ad} \equiv ( \partial ln T / \partial ln P )_S$,
and $\mu$ and $R_g$ are the mean mass per particle and the gas constant.  The
temperature difference at the bottom of the convection zone $\delta T_0$ is
given in terms of parameters $\eta$ and $B_0$ by
\be
\delta T_0 = \eta T_e(0) / \beta_0\ ,
\label{equation:eta}
\ee
where $\beta_0 = 8 \pi P_e / B_0^2$.  The quantity $B_0$ is the magnetic
field strength at the base of the convection zone, and is one of the free
parameters of the problem, while $\eta$ is a dimensionless measure of the
temperature difference at the base of the convection zone, and is constrained
to lie between $0$ and $1$.  When $\eta=0$, the internal temperature is equal
to that of the surrounding plasma, whereas for $\eta=1$, the density inside
the flux tube is equal to that outside.  Together with $L$ and $\Phi$,
specification of $B_0$ and $\eta$ completely defines the flux emergence
problem in our model.  

Values of all these parameters are highly uncertain, so we have attempted to
cover a wide range of parameter space.  For most of our simulations, $B_0$
was chosen to be $10^4{\rm \ G}$, $10^5{\rm \ G}$, or $10^6{\rm \ G}$, and
the flux $\Phi$ was taken as $10^{18}{\rm \ Mx}$, $10^{20}{\rm \ Mx}$,
or $10^{22}{\rm \ Mx}$.  Our choice of these values is discussed further
in Paper I.

\subsection{Example of a Simulation}
\label{subsection:example}

\begin{figure}
\includegraphics[width=5.5in]{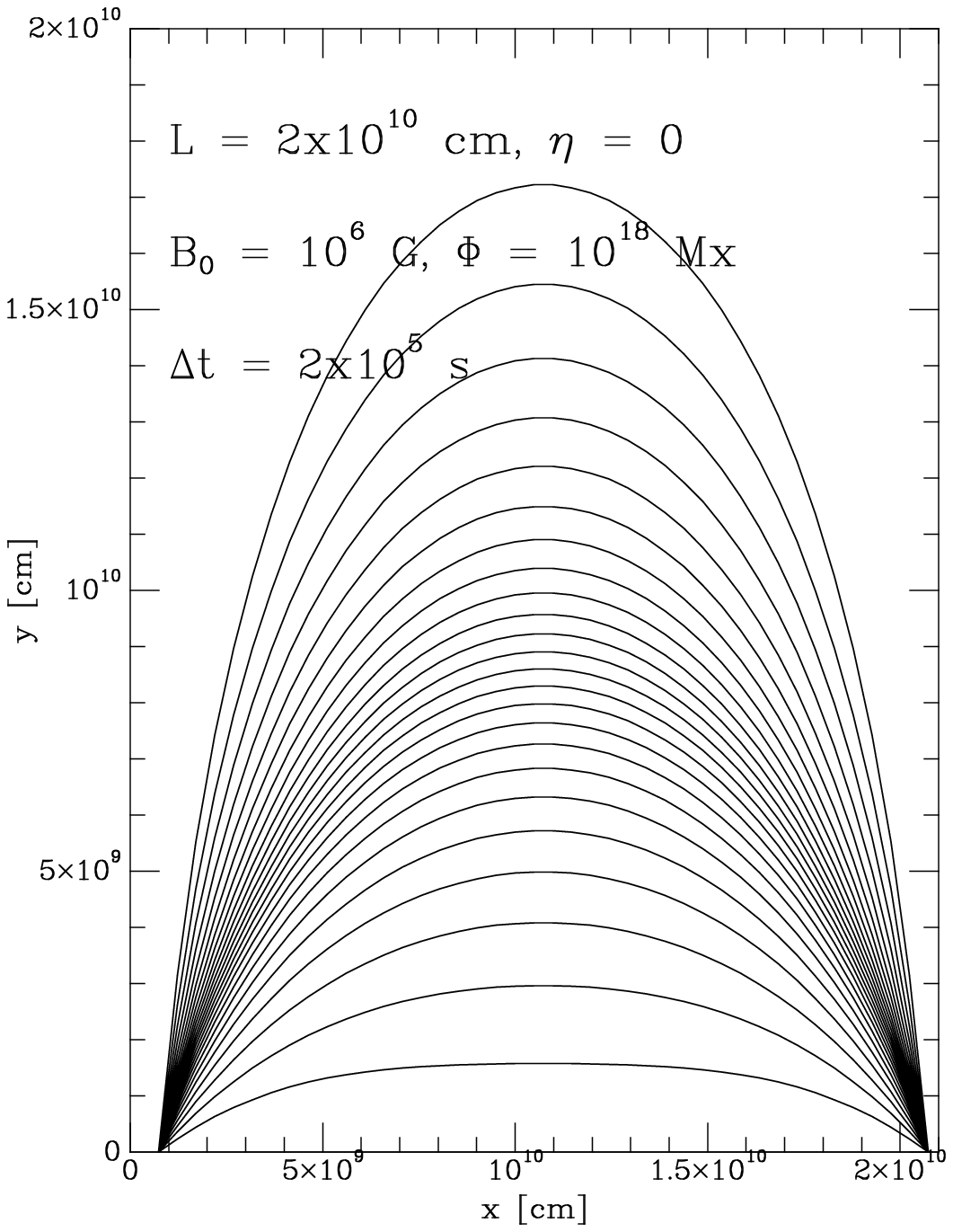}
\caption{
Plot of the flux tube shape at times separated by $2 \times 10^5$ s. 
}
\label{figure:yosfigure1}
\end{figure}

To demonstrate our model, we briefly discuss the results of a single
simulation.  In this case, $B_0 = 10^6{\rm \ G}$, $\Phi = 10^{18}{\rm \ Mx}$,
$\eta=0$, and $L = 2 \times 10^{10}{\rm \ cm}$.  The position of the tube at
numerous times is shown in Figure \ref{figure:yosfigure1}.  This flux tube
is ``unstable'' and emerges from the top of the calculational domain without
reaching an equilibrium configuration.  Note that the upward motion of the
flux tube slows significantly when its apex is roughly half-way through the
convection zone.  This indicates that the magnetic tension was almost strong
enough to balance the buoyancy force.  Had $L$ in fact been only slightly
smaller ($\leq 1.97 \times 10^{10}{\rm \ cm}$), the flux tube would have been
``stable'' and would have stopped rising when the height of the flux tube apex
$y_a \approx 9 \times 10^9{\rm \ cm}$.  We explore this interesting behavioral
dichotomy in the following subsection.

\section{The Critical Length Scale}
\label{section:criticallength}

\begin{figure}
\includegraphics[width=5.5in]{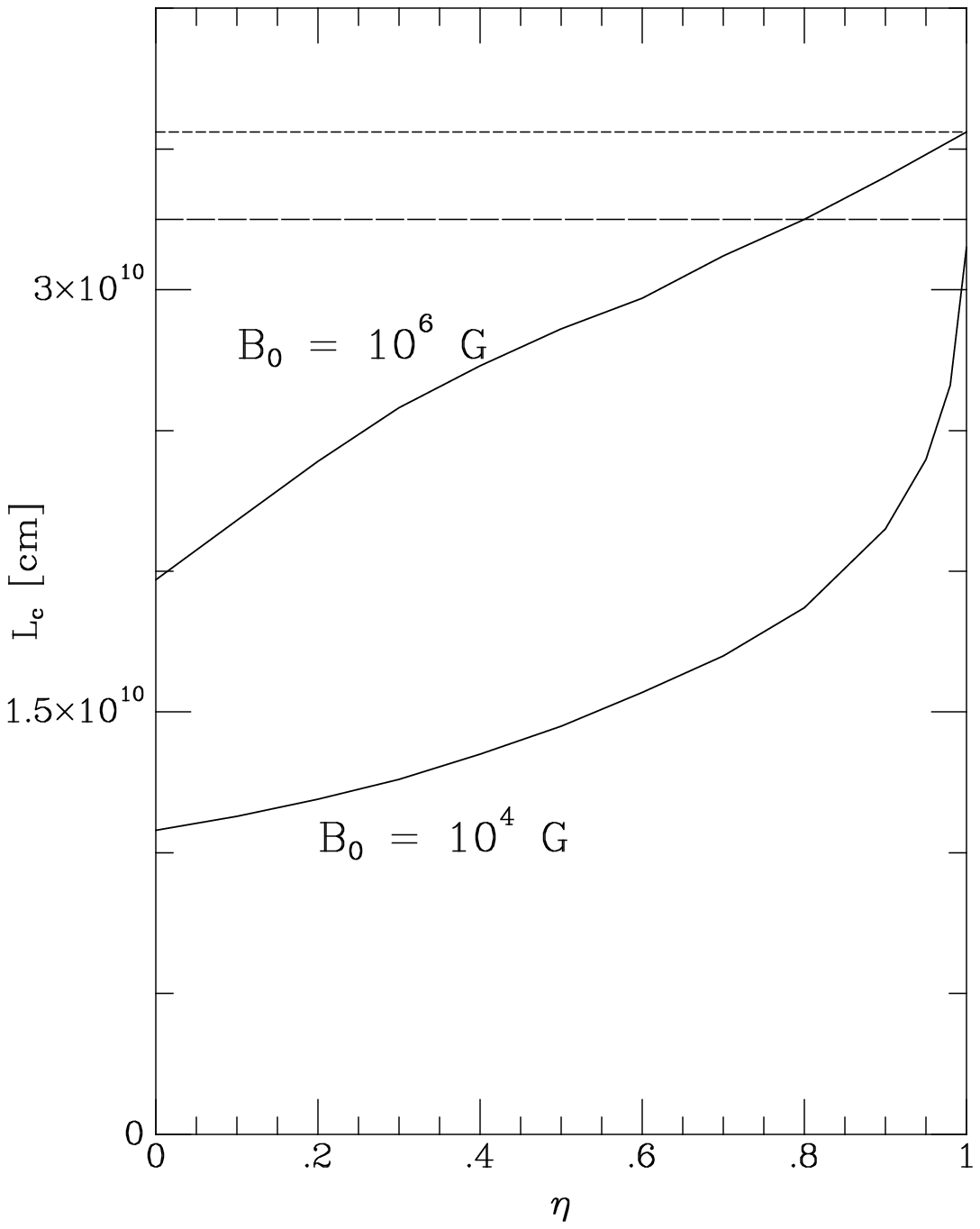}
\caption{
Critical length scale as a function of $\eta$ for $B_0 = 10^6$ and $10^4$ G.
The long-dashed horizontal line corresponds to the value from equation
(\ref{equation:stability}), and the short-dashed line corresponds to
Parker's (1979) isothermal result of $2 \pi \Lambda$.
}
\label{figure:yosfigure2}
\end{figure}

From the discussion in the previous section, it is apparent that for given
values of the other parameters, a critical length $L_c$ exists.  For
$L < L_c$, the flux tube reaches a stable equilibrium, while for $L > L_c$,
an unstable eruption to the photosphere occurs.  Therefore, a knowledge of
the dependence of $L_c$ on the other parameters is very important for 
understanding the nature of flux emergence.  The magnetic flux $\Phi$ affects
the drag force and therefore the rate of rise, but it does not affect the
relative balance between buoyancy and tension, so $L_c$ is independent of
$\Phi$.  To determine the dependence of $L_c$ on $\eta$ and $B_0$, we have
used the numerical model described in the previous section.  In Figure
\ref{figure:yosfigure2}, we plot $L_c$ as a function of $\eta$ for 
$B_0 = 10^4{\rm \ G}$ and $B_0 = 10^6{\rm \ G}$.

There are several features of Figure \ref{figure:yosfigure2} worthy of
mention.  First, note that for fixed $B_0$, $L_c$ increases as $\eta$
increases from $0$ to $1$.  This is easily explained by the decrease in the
relative strength of magnetic buoyancy compared to magnetic tension as the
temperature inside the flux tube decreases.  Thus the footpoint separation
$L$ over which magnetic tension can overcome buoyancy is increased (recall
that for fixed apex height tension will scale roughly as $L^{-2}$).  Second,
as $\eta$ approaches $1$ for fixed $B_0$, $L_c$ approaches a finite value,
even though the buoyant force goes to zero at the base of the convection zone.
Since for $\eta=1$, the initial horizontal state is an equlibrium configuration,
one can perform a linear stability analysis of equations (\ref{equation:eqnmo})
and (\ref{equation:geometry}) (see 
\citealt{SpruitvanB1982},
and Paper II) to obtain $L_c$ analytically
\be
L_c = \sqrt{2/(1 - \nabla )} \pi \Lambda\ ,
\label{equation:stability}
\ee
which is plotted as the horizontal long-dashed line in 
Figure \ref{figure:yosfigure2}.  This analytic result corresponds very well
to our numerical result for $\eta=1$ and $B_0 = 10^4{\rm \ G}$, but it is
about 10\% smaller than the critical length (for emergence of the flux tube
to the photosphere) which we find numerically for $\eta=1$ and 
$B_0 = 10^6{\rm \ G}$.  The reason for this interesting discrepancy is
discussed in Paper II.

\section{Time Scales for Emerging Flux Tubes}
\label{section:timescales}

If magnetic flux emerges from the convective overshoot region and evolves
according to our model, only those flux tubes we have labeled ``unstable''
($L > L_c$) can reach the photosphere.  The time required for the flux tube
to rise through the convection zone is therefore of great interest.  We assume
that a flux tube is formed in the stably stratified region beneath the
convection zone, and begins its buoyant rise when perturbed upward into the
convection zone by gradual heating and convective overshoot motions.
Before perturbation, the flux tube must be in mechanical equilibrium
($\rho_e = \rho_i$) as otherwise it would adjust by rising or sinking
adiabatically 
\citep{Parker1979}.
The nature of a perturbation is described
by the two parameters $L$ and $\eta$.  $L$ is the length scale of the
perturbation, and $\eta$ measures its thermodynamic state.  When $\eta=1$, the
perturbed flux tube is still in mechanical equilibrium ($\rho_i=\rho_e$),
whereas if $\eta=0$, the flux tube is in thermal equilibrium with its
surroundings ($T_i = T_e$).  The proper choice of $\eta$ for the perturbation
can be made by comparing the thermal diffusion time at the bottom of the
convection zone $\tau_{th}$ to the dynamic rise time $\tau_d$ for unstable
$\eta=1$ flux tubes.  If $\tau_{th} >> \tau_d$, then $\eta=1$ is the best
description of the perturbation, whereas if $\tau_{th} << \tau_d$, then
$\eta=0$ is most appropriate.  When the two are comparable, an intermediate
value of $\eta$ is called for.  The thermal diffusion time scale is given by
(see Paper II)
\be
\tau_{th} \approx 0.1 {3 \over 16} {\rho_e^2 C_p a^2 \kappa_R \over 
\sigma T_e^3}\ ,
\label{equation:tauthermal}
\ee
where $\sigma$ is the Stefan-Boltzmann constant, $\kappa_R$ is the Rosseland
mean opacity [${\rm cm}^2{\rm \ g}^{-1}$], and $a=[\Phi/(\pi B_0)]^{1/2}$ is
the radius of the flux tube at the base of the convection zone.  The dynamic
rise time $\tau_d$ for $\eta=1$ can be obtained from the numerical simulations.
Note that the thinner (smaller $\Phi$) flux tubes will have shorter thermal
time scales and longer dynamic time scales.  We find that $\tau_d$ does not
depend strongly on $L$, provided $L$ exceeds $L_c$ by at least 5\%.  We
therefore have used $L=4 \times 10^{10} {\rm \ cm}$ 
(cf. Figure \ref{figure:yosfigure2}) to compute our estimates of $\tau_d$.
Computed values of $\tau_d$ and $\tau_{th}$ as functions of $\Phi$ and $B_0$
are shown in Table 1, 
along with the resultant values of $\eta$.  We also include in Table 1
the ``corrected'' dynamical rise times (denoted $\tau_r$)
using these self-consistent values of $\eta$ in the numerical simulations.

From the information in Table 1 and the critical length
scales of Figure \ref{figure:yosfigure2}, it is now possible to construct a
general picture of flux tube evolution.  For given values of $\Phi$ and $B_0$,
perturbation length scales divide naturally into three ranges: short,
$L < L_c(\eta=0)$; medium, $L_c(\eta=0 ) < L < L_c(\eta = 1)$; and long,
$L_c(\eta = 1) < L$.  If $\tau_{th} << \tau_d$ then only short flux tubes are
stable, while longer flux tubes rise to the photosphere on the dynamical time
scale.  But if $\tau_{th} >> \tau_d$, both short and medium flux tubes are
stable.  Only the short flux tubes are ``absolutely'' stable, however, as the
medium length tubes are able to rise quasi-statically on the thermal time
scale as heat leaks into them.  This is equivalent to gradually decreasing
$\eta$ from $1$ to $0$.  However, before $\eta$ reaches zero, a medium-$L$
flux tube will suddenly find itself with $L > L_c$, and will then erupt to the
photosphere on the much shorter dynamic rise time.

In the most straightforward ($e.g.$ $\alpha \omega$ or $\alpha ^2 \omega$)
kinematic dynamo models the oscillatory nature of the solar cycle is due to
periodic conversion of poloidal to toroidal magnetic field and back via
the mechanisms of stretching by differential rotation and the $\alpha$-effect
(which is usually attributed to net helicity in convective or convective
overshoot motions).  The dynamo period is therefore at least as great as the
time necessary to reorient and stretch poloidal field into toroidal field.  For
these dynamos to work, therefore, magnetic flux must remain in the dynamo
region long enough for reorientation to occur, $i.e.$, for a significant
fraction of the solar cycle.  The conclusions drawn below are based on this
premise.  An alternative picture is that the dynamo generates fields on a much
shorter time scale, in which case our arguments below do not hold.  In that
case, however, the period of the solar cycle itself remains unexplained and
some additional unknown mechanism must be invoked to account for it.

From a comparison of the time scales in Table 1 with the
length of half a solar cycle (roughly 11 years or $3.4 \times 10^8{\rm \ s}$),
some tentative conclusions about the nature of the flux tube perturbations can
be drawn.  First, note that the rise time $\tau_r$ of unstable flux tubes is
always much shorter than the duration of the solar cycle.  If there were a
continuous source of perturbations with length scales $L > L_c$, then magnetic
flux could not remain stably submerged on time scales much longer than $\tau_r$.
One therefore concludes that such perturbations are either rare or nonexistent.
Indeed, if convective overshoot motions are responsible for the flux tube
perturbations, one would expect typical perturbation length scales to be of
order the eddy size, which in mixing length theory is roughly the pressure
scale height near the bottom of the convection zone, 
$i.e.$ $6 \times 10^9{\rm \ cm}$.  This is significantly less than any of the
$L_c$ values in Figure \ref{figure:yosfigure2}.  A perturbation with $L > L_c$
would require coherent action on the part of several adjacent eddies.  Assuming
this to be unlikely, the magnetic flux which does eventually emerge must 
originate from perturbations in the short or intermediate $L$ range.  We
consider two possibilities.

First, if the separation $L$ between anchor points is truly fixed in time, then
only those perturbations in the intermediate $L$ range can emerge at all,
since short $L$ perturbations result in completely stable structures.  
Furthermore, only those combinations of $B_0$ and $\Phi$ which give rise to
$\eta=1$ in Table 1 are viable, $i.e.$ those flux tubes with
$\Phi \approx 10^{22} {\rm \ Mx}$.  In that case, we expect magnetic flux to
emerge to the surface on thermal diffusion time scales and to have a footpoint
separation between $L_c(\eta = 0)$ and $L_c(\eta = 1)$.
\begin{table}
\begin{center}
\caption{Flux Tube Time Scales}
\medskip
\begin{tabular}{lccc}
\tableline\tableline
\  & $B_0 = 10^6$ [G] & $B_0 = 10^5$ [G] & $B_0 = 10^4$ [G]
\\
\tableline
Uncorrected ($\eta=1$) rise time $\tau_d$ [s] & & & \\
\tableline
$\Phi = 10^{22}$ [Mx]  & $8.3 \times 10^{5}$ & 
$6.4 \times 10^{6}$  &
$3.9 \times 10^{7}$ \  \\
$\Phi = 10^{20}$ [Mx] & $1.8 \times 10^{6}$ & 
$1.1 \times 10^{7}$ &
$4.7 \times 10^{7}$ \\
$\Phi = 10^{18}$ [Mx] & $5.3 \times 10^{6}$ & 
$3.0 \times 10^{7}$ &
$8.1 \times 10^{7}$ \\
\tableline
Thermal diffusion time $\tau_{th}$ [s] & & & \\
\tableline
$\Phi = 10^{22}$ [Mx]  & $2.7 \times 10^{7}$ & 
$2.7 \times 10^{8}$ &
$2.7 \times 10^{9}$ \\
$\Phi = 10^{20}$ [Mx]  & $2.7 \times 10^{5}$ & 
$2.7 \times 10^{6}$ &
$2.7 \times 10^{7}$ \\
$\Phi = 10^{18}$ [Mx]  & $2.7 \times 10^{3}$ & 
$2.7 \times 10^{4}$ &
$2.7 \times 10^{5}$ \\
\tableline
Self-consistent $\eta$ & & & \\
\tableline
$\Phi=10^{22}$ [Mx] & 1.0 & 
1.0  &
1.0 \\
$\Phi=10^{20}$ [Mx] & 0.0 & 
0.0  &
0.5 \\
$\Phi=10^{18}$ [Mx] & 0.0 & 
0.0  &
0.0 \\
\tableline
Corrected rise time $\tau_r$ [s] & & & \\
\tableline
$\Phi = 10^{22}$ [Mx] & $8.3 \times 10^{5}$ &
$6.4 \times 10^{6}$ &
$3.9 \times 10^{7}$ \\
$\Phi = 10^{20}$ [Mx] & $6.7 \times 10^{5}$ &
$3.1 \times 10^{6}$ &
$8.6 \times 10^{6}$ \\
$\Phi = 10^{18}$ [Mx] & $2.1 \times 10^{6}$ &
$9.8 \times 10^{6}$ & 
$1.7 \times 10^{7}$ \\
\tableline
\tableline
\end{tabular}
\end{center}
\label{table:table1}
\end{table}

The second possibility is that there is stretching of the anchor point 
separation $L$ past $L_c$ after a stable short $L$ perturbation has been
made.  There are two possible mechanisms for this.  In the first instance,
stretching could be accomplished by differential rotation, for example, if
the anchor points differ in latitude.  The second mechanism is the thermal
heating of the anchor points themselves.  This may cause a gradual
``unzipping'' of the flux tube segment, until the footpoint separation exceeds
$L_c$.  The time scale for this to occur depends on a number of factors, 
including how deeply buried the remainder of the flux tube is in the overshoot
region.  This is discussed further in Paper II.  Here again, for either
mechanism, we expect the anchor point separation of emerging flux to lie in
the range $L_c(\eta = 0) < L < L_c(\eta = 1)$.

As a final point, note that the dispersion in rise times for individual flux
tubes with different values of 
$\Phi$ (a month or more -- see Table 1) is much greater than
the formation times of active regions (typically 2-3 days).  We believe
this indicates that active regions are formed from a single emerging flux tube.
If many tubes of differing sizes were perturbed simultaneously, or if a single
flux tube became highly fragmented near the bottom of the convection zone, one
would expect the emergence of flux to be dispersed over a much greater time
corresponding to the difference in rise times between small $\Phi$ ($e.g.$,
$10^{18}{\rm \ Mx}$) and large $\Phi$ ($e.g.$, $10^{22}{\rm \ Mx}$) flux tubes.
Any fragmentation therefore probably takes place in the topmost portion of the
convection zone when the tube is emerging most rapidly.

\section{Conclusions}
\label{section:conclusions}

We have developed a model, based on the thin flux tube approximation of 
\cite{Spruit1981},
for studying the emergence of magnetic flux through the convection zone
when the footpoints of the flux loop are anchored a distance $L$ apart in
the stable layers below.  Figure \ref{figure:yosfigure1} shows the evolution
of a flux tube whose footpoints are sufficiently far apart ($L > L_c$) to allow
it to rise to the top of the convection zone.  We have used our model to explore
the critical length $L_c$ separating flux tubes which form stable magnetic
loops in the convection zone ($L < L_c$) from those which erupt through the
photosphere ($L > L_c$).  Figure \ref{figure:yosfigure2} shows the dependence
of $L_c$ on the temperature defect parameter $\eta$ for two values of the
field strength at the base of the convection zone $B_0$, $10^4{\rm \ G}$, and
$10^6{\rm \ G}$.  We find that for perturbations on length scales
$L > L_c(B_0,\eta)$ the rise time is short compared to the duration of the
solar cycle.  Under the assumption that the solar cycle can be modeled as
a kinematic ($e.g.$, $\alpha \omega$ or $\alpha^2 \omega$) dynamo model 
operating below the convection zone, we conclude that such perturbations must
therefore be rare or nonexistent.  We speculate that the magnetic flux which
does erupt through the photosphere forms initially from perturbations with
$L < L_c$, resulting in stable structures.  These are subsequently destabilized
either by thermal diffusion or by stretching of the anchor points until
$L$ exceeds $L_c$.  In either case, we expect that the anchor point separation
$L$ should fall roughly within the range $L_c(\eta = 0)$ to $L_c(\eta = 1)$.
The rise time of flux tubes is the time scale for conduction of heat into the
tube in the first case, and the stretching time in the second case.  Finally,
we argue that active regions are formed from the emergence of a single flux
tube segment.

\acknowledgements
G. H. Fisher and A. N. McClymont were supported by NASA under grant NAGW86-4
and by NSF grant ATM-86-19853.  Dean-Yi Chou was supported by the NSC of ROC
under grant NSC 77-0209-M007-01 and by NASA grant NSG-7536 during his time
at the University of Hawaii.  G. H. Fisher was supported by the I.G.P.P. at
Lawrence Livermore National Laboratory for several visits during which some of
this work was completed.


\end{document}